\documentclass[twocolumn]{article}
\usepackage{graphicx} % Required for inserting images
\usepackage{amsmath,amssymb,amsfonts}
\usepackage{algorithmic}
\usepackage{textcomp}
\usepackage{xcolor}
\usepackage{url}
\usepackage{authblk}
\usepackage[sorting = none, backend = bibtex, style=numeric-comp]{biblatex}
\bibliography{Bronnen}
\begin{document}

\renewcommand{\thefootnote}{\fnsymbol{footnote}}

\title{Using continuation methods to analyse the difficulty of problems solved by Ising machines}

\def\correspondingauthor{\footnote{Correspondence and requests should be addressed to J.L. (email: Jacobus.Stijn.T.Lamers@vub.be) or to G.V.d.S. (email: Guy.Van.der.Sande@vub.be)}}

\author[1]{Jacob Lamers \correspondingauthor{}}
\author[]{Guy Verschaffelt}
\author[]{Guy Van der Sande}

\affil[]{Applied Physics Research Group, Vrije Universiteit Brussel, Pleinlaan 2, 1050 Brussels, Belgium}

\date{}

\maketitle

\renewcommand{\thefootnote}{\arabic{footnote}}

\begin{abstract}
Ising machines are dedicated hardware solvers of NP-hard optimization problems. However, they do not always find the most optimal solution. The probability of finding this optimal solution depends on the problem at hand.  Using continuation methods, we show that this is closely linked to the bifurcation sequence of the optimal solution. From this bifurcation analysis, we can determine the effectiveness of solution schemes. Moreover, we find that the proper choice of implementation of the Ising machine can drastically change this bifurcation sequence and therefore vastly increase the probability of finding the optimal solution. 
\end{abstract}

%\begin{IEEEkeywords}
%Ising machines, continuation, nonlinearity
%\end{IEEEkeywords}

\section{Introduction}
In the current era of ever increasing computational demands, the search for energy-efficient computing has become increasingly crucial. Traditional computing architectures often consume substantial amounts of power, leading to significant energy costs and environmental impact \cite{On_global_electricity_usage_of_communication_technology, Global_ICT_emissions, HowToStopDatacentersFromGobblingUpTheWorldsElectricity}. In this context, the Ising machine has emerged as a promising platform that addresses the pressing need for energy-efficient computation. Inspired by the Ising model from statistical physics, Ising machines offer an innovative approach to solve complex combinatorial optimization problems while minimizing energy consumption. 

For many complex optimization problems, it has been demonstrated that the cost function of the problem can be mapped to the energy of an Ising network \cite{Ising_formulations_of_many_NP_problems}. The Hamiltonian of an Ising network describes the interactions between $N$ spins $\sigma_i\in\{-1, 1\}$ with coupling matrix $J_{ij}$ and is given by
\begin{equation}
    \label{eq:Hamiltonian}
    \mathcal{H} = -\frac{1}{2}\sum_{ij}^N J_{ij}\sigma_i \sigma_j.
\end{equation}
The ground state of the Ising network, i.e. set of spins $\vec{\sigma}$ that minimizes the Hamiltonian in equation \eqref{eq:Hamiltonian}, represents the most optimal solution to the problem. %The goal of an Ising machine is therefore to find the ground state of the Ising network corresponding to this optimization problem. An Ising machine is any physical system with the same (or at least similar enough) energy landscape as the target Ising network. This way, after the Ising machine is initialised, it will naturally evolve towards a local energy minimum, preferably the ground state of the Ising network it represents. Of course, a useful Ising machine is able to emulate multiple and preferably all coupling matrices. 
% Zomaar beginnen over Ising network zonder introductie?
%In the Ising machine approach, the optimization problem at hand is first mapped to a network of interacting binary variables, i.e. an Ising network. This gives rise to an energy function of the Ising network that is proportional to the cost function of the optimization problem. 
%connects the cost function of the optimization problem to the energy of the Ising network. 
%The state with the lowest energy, i.e. the ground state, therefore corresponds to the optimum solution. 
This approach of mapping the problem to an Ising Hamiltonian and finding the ground state of this Hamiltonian has already been applied to a multitude of real life problems such as jobs scheduling \cite{Job_Scheduling}, traffic flow regulation \cite{Traffic_Flow}, finance related problems \cite{Fincance} and protein folding \cite{Protein_Folding}. However, it has been shown that finding the ground state of a general Ising network is an NP-hard problem \cite{Computational_complexity_of_Ising_spin_glasses}.

An Ising machine is a piece of hardware dedicated to finding the ground state of a general Ising network. It exploits the tendency of natural systems to move to the lowest energy state. So far, various physical systems have been investigated to act as an Ising machine \cite{OverviewMcMahon} such as superconducting-circuit quantum annealers \cite{Quantum_annealing_DWave} and classical thermal annealers \cite{Conti}.

In this article, we focus on Ising machines that encode the spins in analog variables. These spins' amplitude are left to evolve. Usually, the system ends up in a fixed point where the spin amplitudes no longer change in time. At any point during this evolution, the spin amplitudes can be mapped to binary spins and using equation \eqref{eq:Hamiltonian}, an energy can be associated with the state. This kind of Ising machines are sometimes called dynamical-system solvers \cite{OverviewMcMahon}. These have already been implemented using memristor crossbar arrays \cite{Memristors}, electronic oscillators \cite{electronic_Ising_Machine}, degenerate optical parametric oscillators (DOPO) \cite{DOPO, ChipBasedCIM} and opto-electric oscillators (OEO) \cite{Poor_mans_CIM}. 

In an ideal scenario, at least one of the states visited during the evolution of the Ising machine is the state with the lowest energy. However, this is not always the case, as the system can become trapped in local minima of the Hamiltonian, leading to sub-optimal solutions. Therefore, a lot of research is dedicated to device strategies to improve the probability of reaching this ground state \cite{Bifurcation_behaviors, SimulatedAnnealing, MeanField_Annealing, Optimization_by_mean_field_annealing, Hyperspin_Conti, Scaling_advantage_of_chaotic_amplitude_control, Destabilization_of_local_minima, CIM_ErrorCorrection}. These strategies however treat every problem in the same way. 

Additionally, in Ref.~\cite{Order_of_magnitude}, it was observed that the likelihood of reaching the ground state heavily depends on the specific problem and the physical implementation of the Ising machine. Therefore, it is crucial to investigate why some problems are more challenging than others and understand the impact of the Ising machine's implementation. Such insights are vital for comprehending the functionality of the Ising machine and enhancing its performance. 

To address these questions, we employ continuation methods on various benchmark problems. We investigate the bifurcation sequence of the optimum solution. We show how this is key to characterise the difficulty of the optimization problem at hand. We also demonstrate that this bifurcation sequence depends on the physical implementation of the Ising machine. Therefore properly designing the Ising machine can render some problems easier to solve. 

The remainder of the paper is structured as follows. In section \ref{sec:Difficulty_Classes}, we use continuation techniques to analyse the difficulty of finding the ground state of Ising networks and introduce three difficulty classes. In section \ref{sec:Changing_difficulty_class}, we determine the difficulty class of a problem when considering two different physical implementation of the Ising machine. To gain more insight in the previous observations, we focus in section \ref{sec:4Spins} on the bifurcation sequence of a four spin toy model. Finally, in section \ref{sec:Problem_g05_100_1}, we determine the difficulty class of all unweighted benchmark problems of the BiqMac library when using three different physical implementations of the Ising machine. We conclude that the insights gained in section \ref{sec:4Spins} are still valid for larger size problems.

\section{Difficulty classes when using Ising machines}
\label{sec:Difficulty_Classes}
The spins of the Ising network are implemented using bistable gain-dissipative systems. The two stable states of this system represent the spin-up and spin-down state. These bistable systems can be coupled to each other in the same way as the corresponding spins in the Ising network. The analog variables that are used to emulate the binary spins are referred to as spin amplitudes. A general set of dimensionless equations modelling the evolution of these spin amplitudes of the Ising machine $x_i$ is given by 
\begin{equation}
    \label{eq:GeneralDynamics}
    \frac{dx_i}{dt} = \mathcal{F}_i\left(x_i, \alpha, \beta, \sum_j^N J_{ij}x_j\right),
\end{equation}
where $\mathcal{F}$ is the nonlinear function giving rise to the bistability of the gain-dissipative system and $\alpha$ is the linear gain. The spins are coupled using a coupling strength $\beta$. For simplicity, we ignore the time of flight which might exist in the coupling. In principle, a noise term should be added to equation \eqref{eq:GeneralDynamics}, but as we are focusing on bifurcation analysis with continuation methods, this noise is set to zero. At any point during the evolution of the spin amplitudes, the binary spins can be obtained by taking the sign of the spin amplitudes
\begin{equation}
    \label{eq:map}
    \sigma_i = \text{sign}\left(x_i\right).
\end{equation} 
Remark that the Ising machine energy is calculated from equation \eqref{eq:Hamiltonian} only after applying the mapping of equation \eqref{eq:map}.

We will first focus on a pitchfork normal form as a transfer function. The pitchfork normal form occurs naturally in various optical systems and has been used to describe Ising machines for e.g. the classic approximation of the coherent Ising machine \cite{CIM, DOPO} and polariton condensates \cite{PolaritonCondensates}. In this case, the set of dynamical eqations now reads:
\begin{equation}
    \label{eq:ThirdOrder}
    \frac{dx_i}{dt} = (\alpha-1)x_i - x_i^3 + \beta\sum_j^N J_{ij}x_j.
\end{equation}

MaxCut is the task of dividing a network in two sub-networks such that the number of connections between these two sub-networks is maximized. We chose this task as it can be mapped to an Ising network in a straightforward way. We focus on the random, unweighted graphs with $50\%$ edge probability contained in the BiqMac library \cite{BiqMac}. For these problems, the optimal solution is known. We have applied continuation algorithms to these benchmark tasks using the software 'auto-07p' \cite{AUTO}. These algorithms take a fixed point as an input and can then follow that state in phasespace as the parameters $\alpha$ and $\beta$ are changed. This can be done for both stable and unstable fixed points. Furthermore, bifurcation points are detected and the algorithm can also track new branches of fixed points originating at such a bifurcation point. %We use these continuation tools to track both the fixed point with the ground state energy and bifurcations occurring at the origin. This way, we investigate when the latter becomes stable or bifurcates and whether or not it is connected in some way to the former. 

As an illustration, we focus first on three MaxCut problems. A first problem consists of $100$ spins arranged in a $10\times10$ grid with antiferromagnetic interactions between nearest neighbors with periodic boundary conditions. The two other problems are g05\_100.1 and g05\_100.2 from the BiqMac library. 

We first discuss the continuation analysis of the $10\times 10$ grid problem, which is shown in figure \ref{fig:Difficulty_Classes}~(a). In this figure, we initialised the continuation at a $\beta$-value of zero and with all spin amplitudes equal to zero. This is a trivial fixed point of the system, which we will refer to as the origin. This origin exists for all values of $\beta$. We track this state from $\beta=0$ to higher values of the coupling strength $\beta$ and find that for small $\beta$-values ($\beta < 0.25$), it is the only stable state of the system and is unstable for $\beta > 0.25$. For clarity, this state is not shown in figure \ref{fig:Difficulty_Classes}~(a) for $\beta > 0.25$. At $\beta=0.25$, two new stable fixed points emerge at the origin. They are related to each other by the symmetry operation that flips the sign of all spins and therefore have the same energy. The change in stability of the fixed points at $\beta=0.25$ is called a pitchfork bifurcation (PB). The collection of fixed points that is obtained by tracking a state through parameter space is referred to as a branch. We call these two stable branches that emerge at the first PB of the origin the first pitchfork branches. Because of the symmetry, any change in stability that occurs on one first pitchfork branch, will occur on the other one. Therefore, for clarity, in figure \ref{fig:Difficulty_Classes}~(a), only one of these branches is shown. In this figure, all spin amplitudes of the first pitchfork branch have the same absolute value, which increases as $\beta$ is increased. Remark that the spin amplitudes of all $100$ spins of the problem are shown, some with positive sign, some with negative sign. As they all have the same absolute value, the lines overlap. For this problem, the energy of the first pitchfork branch is, for all $\beta$ where it exists, equal to the ground state energy. This is indicated by the gray background. Near the pitchfork bifurcation at the origin, the evolution of the spin amplitudes is determined by the principal eigenvector of the coupling matrix. After this bifurcation, no spin amplitudes of the branch change sign. Therefore the ground state was already given by the sign of the components of this principal eigenvector. This is why such problems are referred to as "spectral easy", adopting the terminology from \cite{Ganguli_conference}. It is well known that these problems can be solved in polynomial time \cite{Berloff}. Also note that, as $\beta$ is increased, plenty more branches will emerge and become stable. However, these are not shown in figure \ref{fig:Difficulty_Classes} as we are mainly interested in the emergence and change of stability of the ground state, i.e.~the bifurcation sequence of the ground state.

A different scenario unfolds in the second exemplary problem, shown in figure \ref{fig:Difficulty_Classes}~(b). We show the results of the continuation analysis of problem g05\_100.2 of the BiqMac library. Again, the continuation was started from $\beta=0$ with all spin amplitudes equal to zero and we track this state to higher $\beta$-values. This time, the origin becomes unstable through a PB at $\beta=0.10$ and again, we do not plot the origin for higher $\beta$-values in figure \ref{fig:Difficulty_Classes}~(b). At the PB, again two first pitchfork branches emerge, only one of which is shown in the figure. In this case, the first pitchfork branch does not have the ground state energy for $\beta$-values immediately after the PB. Note that the spin amplitudes of the first pitchfork branch follow the principal eigenvector of the coupling matrix for $\beta$-values close to the PB. However, as $\beta$ is increased, the steady state deviates more and more from that eigenvector. We even observe that as $\beta$ increases, some spin amplitudes change signs, corresponding to binary spin flips. Because of these spin flips, the steady states along the branch shown in figure \ref{fig:Difficulty_Classes}~(b) do not represent the same binary spin configuration and energy for all $\beta$-values, when using the mapping in equation \eqref{eq:map}. It is only for $\beta$-values higher than $0.242$, which is indicated by the gray background in figure \ref{fig:Difficulty_Classes}~(b), that the sign of the values of the spin amplitudes corresponds to the ground state. Unlike spectral easy problems, the ground state of these problems cannot be obtained immediately after the pitchfork bifurcation and is thus different from the sign of the components of the principal eigenvector of the Jacobian matrix at the origin. The exact value of the minimal coupling strength at which the first pitchfork branch gets mapped to the ground state by applying equation \eqref{eq:map} is initially unknown. We classify such problems as "Ising easy" \cite{Ganguli_conference}.

Compared to figure \ref{fig:Difficulty_Classes}~(a), it is clear that the spin amplitudes are now far more inhomogeneous. Inhomogeneity of spin amplitudes has already been discussed in the literature \cite{Order_of_magnitude, CombinatorialOptimization_Leleu}. When the spin amplitudes do not all have the same absolute value, it is not guaranteed that applying the mapping in equation \eqref{eq:map} on the spin amplitudes will deliver a minimum of the binary Ising system. This is only assured in the limit of $\beta$ going to infinity \cite{paper_Ganguli}. This means that $\beta$ has to be large enough before the correct signs are obtained by the Ising machine. 

Ising easy problems can be solved efficiently using so called annealing schemes. The system is then integrated using equation \eqref{eq:ThirdOrder} with parameter-values just above the first pitchfork bifurcation of the origin ($\beta>0.1$ in figure \ref{fig:Difficulty_Classes}~(b)) from non-zero initial conditions. Given enough time, this integration will lead to one of the first pitchfork branches. Then $\beta$ is increased slowly enough so that the simulation will naturally relax to the first pitchfork branch \cite{paper_Ganguli, 100000SpinsCIM, CIM_with_QuanumAdiabaticAnnealing}. In the case of Ising easy problems, when $\beta$ is sufficiently high, the system inevitably reaches the ground state. We show in the supplementary material note \ref{sec:Annealing} that noiseless simulations of the Ising machine can find the ground state of Ising easy problems with $100\%$ success rate.

\begin{figure}
     \centering
     \includegraphics[width=.99\linewidth]{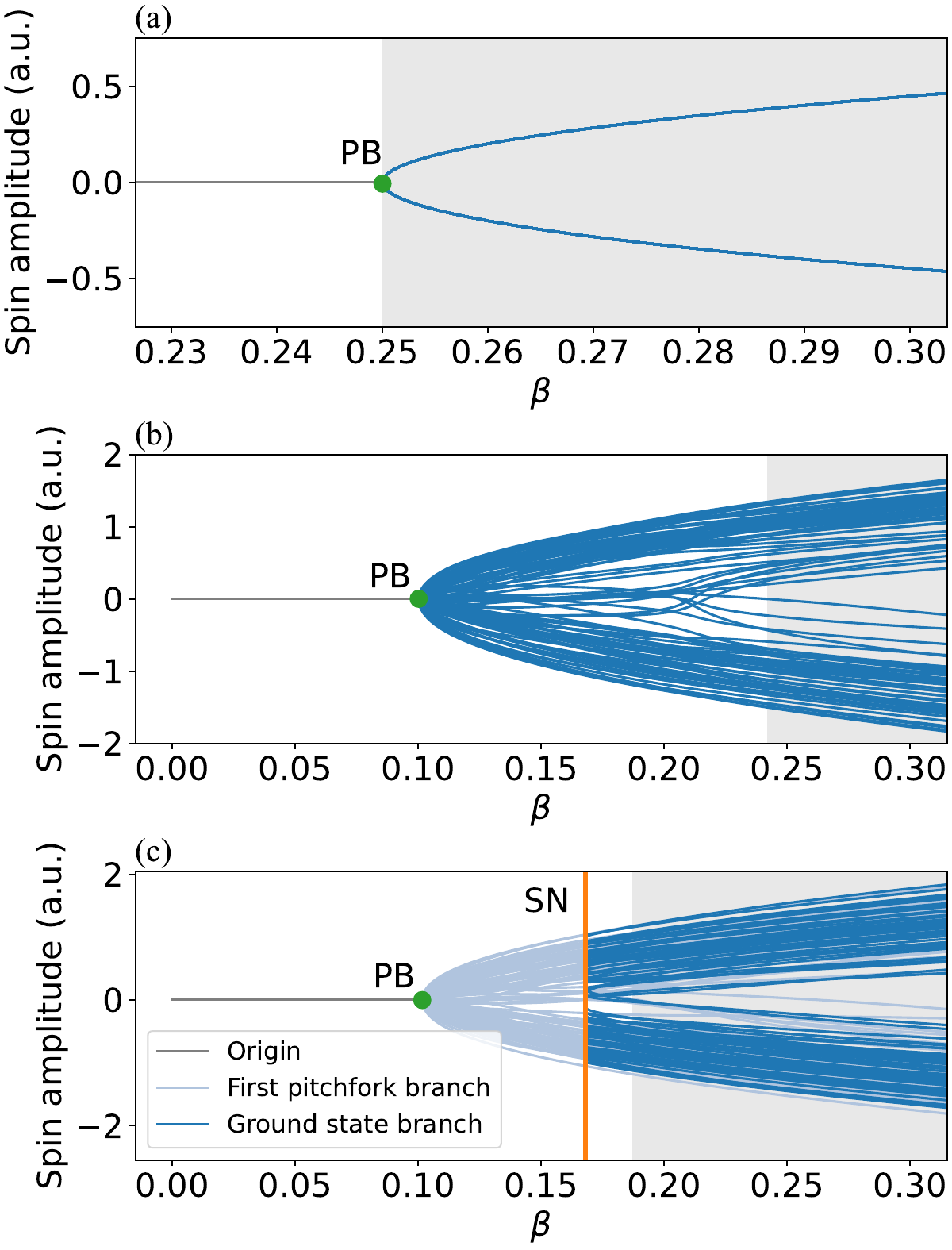}
     \caption{Spin amplitudes while tracking the ground state using continuation when using equation \eqref{eq:ThirdOrder} with $\alpha=0$. (a) is the result of a problem with $100$ spins and antiferromagnetic, nearest neighbours interactions, (b) of problem g05\_100.2 of the BiqMac library and (c) of problem g05\_100.1.}
     \label{fig:Difficulty_Classes}
\end{figure}

As an example of a third type of problem, we show in figure \ref{fig:Difficulty_Classes}~(c) the results of the continuation analysis of problem g05\_100.1 of the BiqMac library. If we follow the same approach as before, we find a similar scenario as in figure  \ref{fig:Difficulty_Classes}~(b). The origin is destabilized at a PB at $\beta=0.10$. At that PB, the first pitchfork branches emerge. However, contrary to what happens for Ising easy problem, the energy of the first pitchfork branch never reaches the ground state energy for any value of $\beta$. Therefore this branch is indicated in light blue in figure \ref{fig:Difficulty_Classes}~(c). Nevertheless, we have obtained the ground state using numerical simulations of equation \eqref{eq:ThirdOrder} (see Methods) at $\alpha=0$ and $\beta=1$. We insert this solution into the continuation tool and use it to track this fixed point back to lower values of $\beta$. In this way, we obtain a branch of fixed points to which we will refer as the ground state branch, which we have plotted in figure \ref{fig:Difficulty_Classes}~(c) in dark blue.  This ground state branch is stable and disappears when the coupling strength $\beta$ is decreased below $0.168$. This is indicated by the orange line in figure \ref{fig:Difficulty_Classes}~(c). This bifurcation point is called a saddle node bifurcation (SN). 

In order to show more clearly what happens in this scenario, in figure \ref{fig:g05_100.1_Spin3}, we only plot the value of spin amplitude $3$. In this figure, solid lines correspond to stable states and dashed lines to unstable states. Again, at low values of $\beta$, the origin is stable and it destabilizes at the first PB indicated by a green dot. At this PB, the first pitchfork branches emerge from the origin, only one of which is shown. The ground state branch and an unstable branch, emerge from the SN and these branches never connect to the first pitchfork branch. Remark that it is the full state that bifurcates and that you would see a similar figure for all the other spins. Such problems, where the first pitchfork branch never maps to the ground state energy are classified as 'Ising hard' problems \cite{Ganguli_conference}.

Due to this gap between the ground state branch and the first pitchfork branch, annealing schemes are not effective in solving Ising hard problems. When using annealing schemes, the Ising machine would follow the first pitchfork branch and will therefore never have the ground state energy. In order to bridge the gap between the first pitchfork branch and the ground state branch, one would have to rely on noise. However, the system has in general a lot of other stable fixed points that are not shown in figure \ref{fig:g05_100.1_Spin3}. So the chance of jumping to the ground state branch due to noise will in general be very low. Furthermore, there is no way to know in advance the $\beta$-value for which the SN happens. This is why Ising hard problems are generally hard to solve.

In figures \ref{fig:Difficulty_Classes} and \ref{fig:g05_100.1_Spin3}, the plots are obtained using $\alpha=0$. However, using different values of $\alpha$ only shifts the bifurcation points but does not change the bifurcation sequence. The underlying reason is discussed in the supplementary material \ref{subsec:Renormalisation}.

When considering the bifurcation sequence of the ground state branch, there are three possibilities. The ground state branch can either be directly connected to the first PB bifurcation of the origin or not. When it is directly connected, it can either immediately attain the ground state energy or $\beta$ has to be increased before it does so. So in this regard, all Ising problems belong to one of the difficulty classes that are discussed above. There are effective methods to tackle both spectral and Ising easy problems which will have a large probability to find the ground state. Ising hard problems can be much harder to tackle and therefore, the difficulty classes discussed above can explain why some problems are more difficult to solve using Ising machines than others.

\begin{figure}
    \centering
    \includegraphics[width=0.99\linewidth]{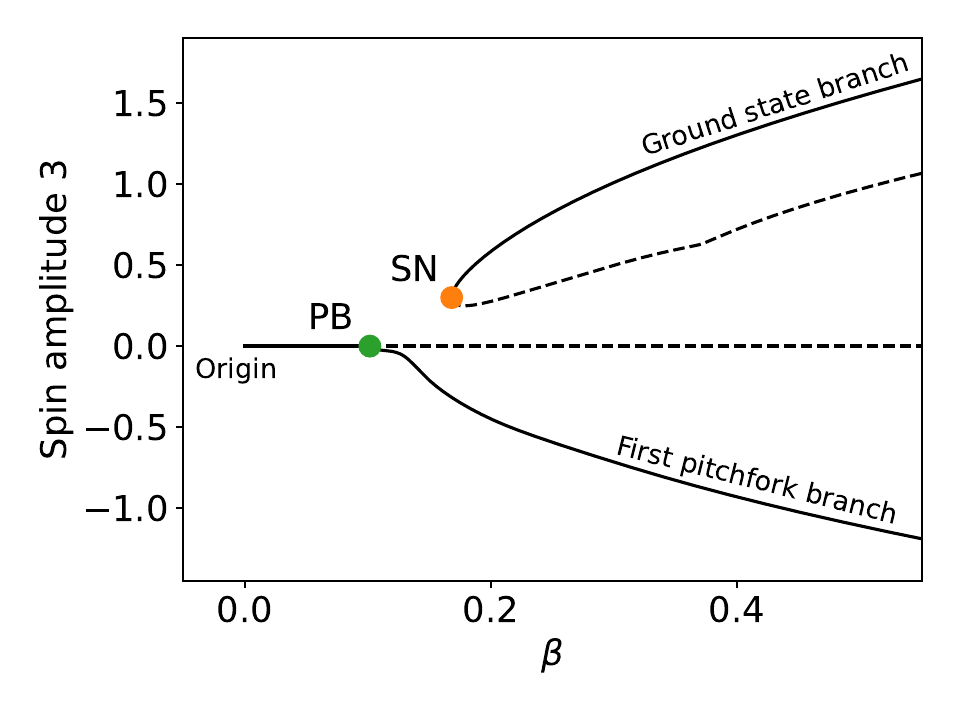}
    \caption{Continuation of spin amplitude 3 of problem g05\_100.1 when solved using the third order nonlinearity ($\alpha=0$). Solid (dashed) lines are used when this spin amplitude is part of a  stable (unstable) state.}
    \label{fig:g05_100.1_Spin3}
\end{figure}

\section{Influence of altering the nonlinearity on the difficulty class of an Ising problem}
\label{sec:Changing_difficulty_class}

In \cite{Order_of_magnitude}, it was observed that using a different physical implementation, which comes down to using a different nonlinearity, can also heavily influence the success rate of the Ising machine in solving a particular problem. As we observe that the bifurcation sequence of the ground state defines the difficulty class, we conjecture that the nonlinearity will influence this sequence in a nontrivial way. We therefore investigate this influence of the nonlinearity on the difficulty classes. We first focus on the sigmoid nonlinearity. This particular nonlinearity can be utilized e.g. to account for the saturation of the spin amplitude commonly observed in experimental setups. The equation describing the dynamic evolution of the analog spin amplitudes $x_i$ is now given by
\begin{equation}
    \label{eq:tanh}
    \frac{dx_i}{dt} = -x_i + \tanh{\left(\alpha x_i + \beta\sum_j^N J_{ij}x_j\right)}.
\end{equation}
Note that (see also supplementary material), contrary to the third order polynomial nonlinearity in equation \eqref{eq:ThirdOrder}, we now have two effective parameters: the linear gain $\alpha$ and the coupling strength $\beta$. 

We apply the continuation analysis using this sigmoid nonlinearity to the Ising hard problem that was studied in the previous section, i.e. problem g05\_100.1 from the BiqMac library. The results are shown in figure \ref{fig:tanh_different_alpha} for two different values of $\alpha$. In figure \ref{fig:tanh_different_alpha}~(a), the continuation performed using equation \eqref{eq:tanh} is done using $\alpha=0.98$. Once again, for low values of the coupling strength $\beta$, the origin is the only stable state of the system. This state becomes unstable at the first PB of the origin (at $\beta=0.00203$) and at this bifurcation point the first pitchfork branch emerges. When the third order polynomial nonlinearity was used to solve this problem, this first pitchfork branch was stable for all $\beta$-values larger than the $\beta$-value of the PB. Here, when using the sigmoid nonlinearity, the first pitchfork branch disappears through a SN (SN 1 in figure \ref{fig:tanh_different_alpha}~(a)). The unstable branch of that SN loops back to lower values of the coupling strength $\beta$ and connects to the ground state branch via a second SN (SN 2 in figure \ref{fig:tanh_different_alpha}~(a)). The energy of this branch is immediately the ground state energy, which is indicated by the gray background in figure \ref{fig:tanh_different_alpha}~(a).

Next, we change the value of the gain $\alpha$ in equation \eqref{eq:tanh} to $\alpha=0.95$ and investigate what changes in the bifurcation sequence. This continuation analysis is shown in figure \ref{fig:tanh_different_alpha}~(b). Contrary to the case with $\alpha=0.98$, the energy of the first pitchfork branch does reach the ground state energy for $\beta>0.023$. The SNs are no longer present and the ground state can now be obtained with the use of annealing schemes. Therefore, when using the sigmoid nonlinearity, problem g05\_100.1 is Ising hard for $\alpha=0.98$, but becomes Ising easy for $\alpha=0.95$. So, choosing the gain parameter $\alpha$ appropriately can alter the difficulty class of a problem when using the sigmoid nonlinearity.

\begin{figure}
     \centering
     \includegraphics[width=.99\linewidth]{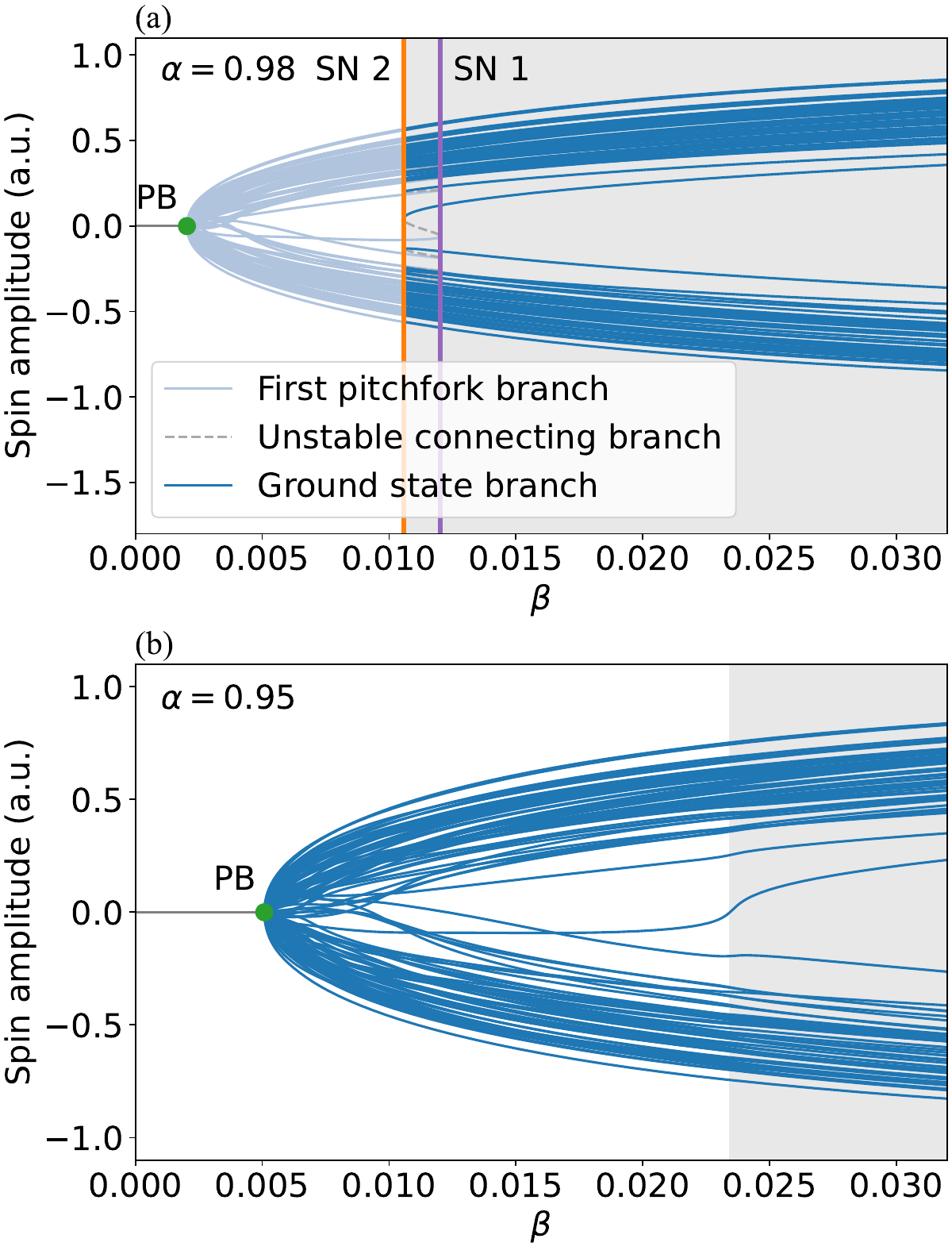}
     \caption{Continuation of the spin amplitudes of the first pitchfork branch and the ground state branch of problem g05\_100.1 of the BiqMac library using the sigmoid nonlinearity with (a) $\alpha=0.98$ and (b) $\alpha = 0.95$. They gray shaded background indicates the region where the energy of the ground state branch is the ground state energy.}
     \label{fig:tanh_different_alpha}
\end{figure}

\section{Four spin toy model}
\label{sec:4Spins}
In order to investigate and better understand why some problems can switch difficulty class when solved with the sigmoid nonlinearity as described in the previous section, we take a closer look at a small problem consisting of four spins. The interactions between these four spins are encoded in the following coupling matrix
\begin{equation}
    \label{eq:Coupling_N4}
    J=\begin{pmatrix}
        0&0.2608&-0.54&-0.2608\\
        0.2608&0&0.1992&-1\\
        -0.54&0.1992&0&-0.1992\\
        -0.2608&-1&-0.1992&0
    \end{pmatrix}.
\end{equation}
This coupling matrix was chosen solely because it is an Ising hard problem when solved with the sigmoid nonlinearity when using an $\alpha$ larger than $0.9919$. Since this is a small problem, the ground state can be found by going over all possible spin states and it turns out to be $\Vec{\sigma}_{\text{GS}} = [1,1,-1,-1]$.

%This time, we use the software 'auto-07p' to perform the continuations \cite{AUTO}. All continuations in the remaining of this article will be performed with this software.

Figure \ref{fig:Spins_VS_beta_tanh} shows the spin amplitudes of the first pitchfork branch as a function of the coupling strength $\beta$. This continuation was performed using the sigmoid nonlinearity in equation \eqref{eq:tanh} and a fixed $\alpha=0.996$. The first pitchfork branch emerges through a PB at $\beta=0.00353$. Above, but close to this bifurcation, the dynamical behavior of the spin amplitudes is governed by the principal eigenvector of the coupling matrix. % Since this Jacobian matrix and the coupling matrix only differ by a constant diagonal matrix, they commute and have the same eigenvectors. 
The sign of the components of this principal eigenvector are $\Vec{u}_1 = [1, 1, 1, -1]$. In accordance with this eigenvector, we see in figure \ref{fig:Spins_VS_beta_tanh} that three of the spin amplitudes take on positive values and the fourth spin amplitude becomes negative after the PB. This means that, starting from the first pitchfork branch, the third spin amplitude has to change sign for the Ising machine to obtain the ground state. The first pitchfork branch disappears through SN 1 and the unstable branch of SN 1 loops back to lower values of $\beta$ and connects to a different branch through SN 2. This branch is the ground state branch as the third spin amplitude changed signs on the unstable connecting branch. 

Note that this problem (for this value of $\alpha$) is Ising hard and cannot be tackled efficiently by annealing schemes. The annealing schemes are designed to track the first pitchfork branch as $\beta$ is increased. As this branch ceases to exists after SN 1, the Ising machine will then be attracted to a different stable fixed point and for a general problem, there will be no guarantee that this will be the ground state branch.

\begin{figure}
     \centering
     \includegraphics[width=\linewidth]{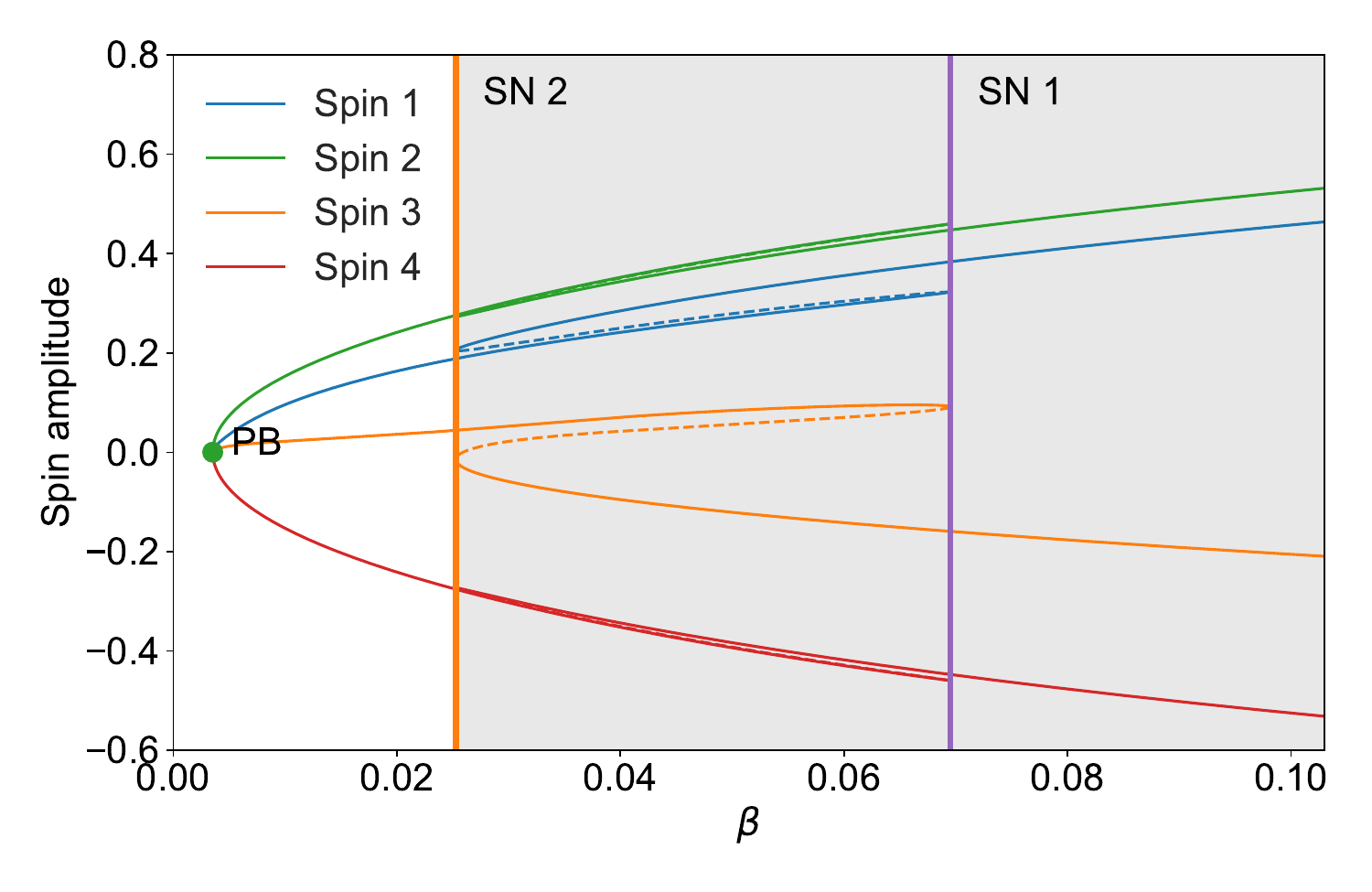}
     \caption{Continuation of the spin amplitude of all four spins of the first pitchfork branch. This continuation was performed using the sigmoid nonlinearity for fixed $\alpha=0.996$. The lines are solid when the corresponding fixed point is stable and dashed when the corresponding fixed point is unstable. They gray shaded background indicates the region where the energy of the ground state branch is the ground state energy.}
     \label{fig:Spins_VS_beta_tanh}
\end{figure}

\begin{figure}
     \centering
     \includegraphics[width=\linewidth]{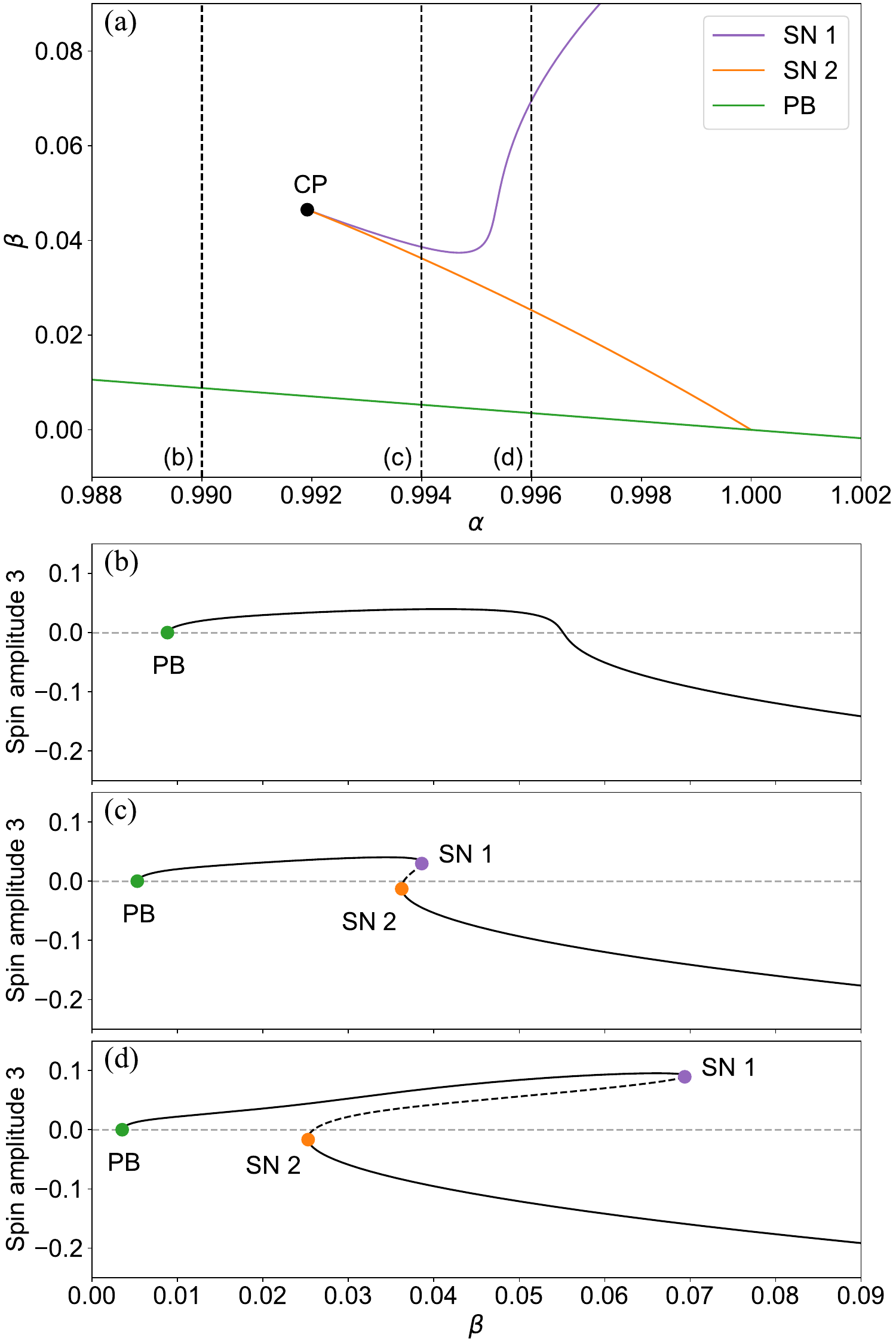}
     \caption{(a) Continuation in the two parameters ($\alpha$ and $\beta$) of the pitchfork bifurcation point (PB) and the two saddle node bifurcations (SN) for the four-spin-system solved with the sigmoid nonlinearity in equation \eqref{eq:tanh}. The three dashed lines represent the three $\alpha$-values at which the other three plots are obtained. The value of the third spin amplitude as a function of the coupling strength $\beta$ is shown in (b) for $\alpha=0.990$, in (c) for $\alpha=0.994$ and in (d) for $\alpha=0.996$.}
     \label{fig:3Regions_Cusp}
\end{figure}

So for $\alpha=0.996$, the ground state branch is connected to the PB via two SNs. We can now use continuation tools to track these three bifurcations through the parameter space of $\alpha$ and $\beta$. Figure \ref{fig:3Regions_Cusp}~(a) shows the position of the two saddle node bifurcations (SN 1 and SN 2) and of the PB for different values of $\alpha$ and $\beta$. In order to further illustrate the meaning of this figure, three cuts at three different values of $\alpha$, indicated by the dashed lines in \ref{fig:3Regions_Cusp}~(a), are shown in figures \ref{fig:3Regions_Cusp}~(b)-(d). These figures show the amplitude of spin amplitude 3 as a function of $\beta$. This spin amplitude is chosen as it is the only one that needs to switch its sign to reach the ground state. Remark that it is the full state that bifurcates, not the single spin amplitude itself.

Figure \ref{fig:3Regions_Cusp}~(d) is made using $\alpha=0.996$, which corresponds to the situation shown in figure \ref{fig:Spins_VS_beta_tanh}. As can be seen in figures \ref{fig:3Regions_Cusp}~(a) and (c), the two SNs (SN1 and SN 2) move closer and closer together as $\alpha$ is decreased. When the gain $\alpha$ is decreased further, the two saddle node bifurcations collide in a cusp point (CP) at $\alpha_{\text{CP}} = 0.9919$. For lower $\alpha$-values, the two SNs have disappeared. In this case, annealing schemes can be used to obtain the ground state as the fixed point with the ground state energy is now connected directly to the first PB of the origin, as is shown in figure \ref{fig:3Regions_Cusp}~(b). This means the problem is Ising easy for $\alpha<\alpha_{\text{CP}}$.

When we used the third order polynomial nonlinearity in section \ref{sec:Difficulty_Classes}, we could not change the difficulty class of a problem by changing the Ising machine's parameters. This is because there is only one effective parameter that can be tuned in this model. As the difficulty classes are defined by the bifurcation sequence with respect to one parameter, a second free parameter is required to change this sequence. The sigmoid nonlinearity in equation \eqref{eq:tanh} does have a second independent parameter which can be used to change the difficulty class of a problem. One can now wonder if adding a second independent parameter to the third order polynomial nonlinearity is enough to enable the same ability. As the third order polynomial nonlinearity can be seen as the first three orders of a Taylor expansion of the sigmoid nonlinearity, a natural way of adding a second parameter is to add the next contribution to this Taylor expansion. The dynamical equations describing the analog spin amplitudes are then given by
\begin{equation}
    \frac{dx_i}{dt} = (\alpha-1)x_i - x_i^3 - \zeta x_i^5 + \beta\sum_j J_{ij}x_j,
    \label{eq:FifthOrder}
\end{equation}
where $\zeta$ is the fifth order parameter. These equations can e.g. be used to model the dynamics of Kerr-nonlinear microring resonators \cite{MicroringResonators}. Similar to the case for the third order polynomial nonlinearity, the gain parameter $\alpha$ can be set to zero without loss of generality by a rescaling of the time and the spin amplitudes. Also note that the third order polynomial nonlinearity is retrieved when $\zeta\rightarrow 0$. Using continuation tools on the origin when using the transfer function in equation \eqref{eq:FifthOrder} results in a plot (not shown here) that is qualitatively similar to figure \ref{fig:Spins_VS_beta_tanh}. Figure \ref{fig:3Regions_Cusp_5thOrder} shows the positions of both saddle node bifurcations as well as the pitchfork bifurcation for various values of the parameters $\beta$ and $\zeta$. As usual the PB is encountered at low values of $\beta$. This bifurcation is independent of $\zeta$. The position of SN 1 and SN 2 however, strongly depend on the value of $\zeta$. For $\zeta$-values below that of the cusp point (CP) at $\zeta = 0.017$, these two SNs exist indicating again that the first pitchfork branch is only connected to the ground state branch via an unstable branch. In this region, the problem is thus Ising hard. When $\zeta$ is decreased, these two SNs move further away from each other. The $\beta$-value of SN 1 tends to infinity as $\zeta$ tends to zero. This means that when using the third order polynomial nonlinearity ($\zeta=0$), the first pitchfork branch remains stable for all values of $\beta$. The ground state still disappears through SN 2 when $\beta$ is decreased, but the unstable branch only connects to the first pitchfork branch at $\beta = \infty$. When $\zeta$ is larger than its value at the CP, the SNs have disappeared and the ground state branch is directly connected to the first pitchfork branch. Therefore, the fifth order polynomial nonlinearity is able to render the problem Ising easy for $\zeta > 0.017$.

With this continuation analysis on a simple model, we thus see how Ising machine implementations has a profound influence on the bifurcation sequence. It also explains why some implementations have higher performance as they can render Ising hard problems into Ising easy ones.

\begin{figure}
     \centering
     \includegraphics[width=\linewidth]{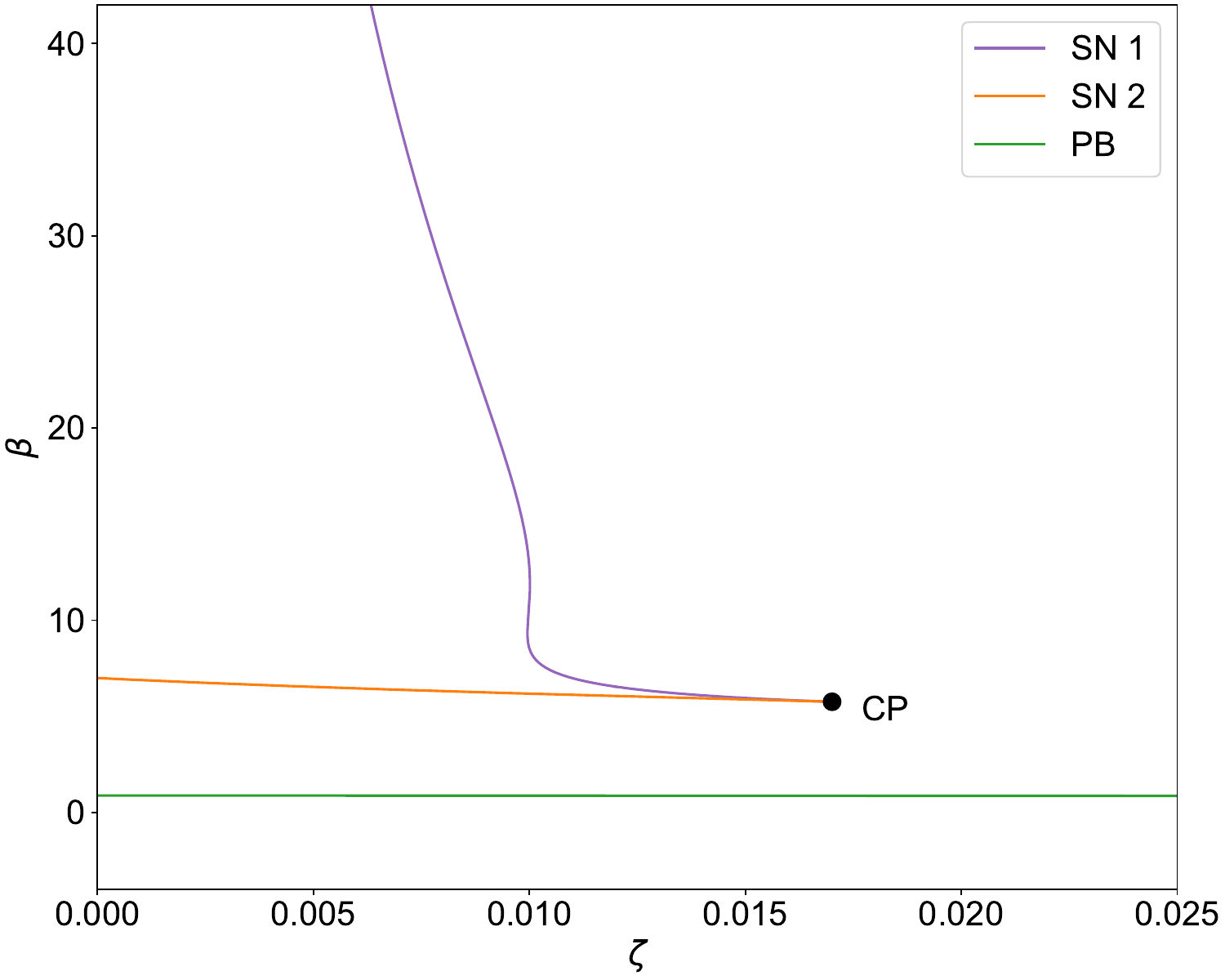}
     \caption{(a) Continuation in the two parameters ($\zeta$ and $\beta$) of the pitchfork bifurcation point (PB) and the two saddle node bifurcations (SN) for the four-spin-system solved with the 5th order polynomial nonlinearity in equation \eqref{eq:FifthOrder}. The value of the third spin amplitude as a function of the coupling strength $\beta$ is shown in (b) for $\zeta=0.007$, in (c) for $\zeta=0.012$ and in (d) for $\zeta=0.020$.}
     \label{fig:3Regions_Cusp_5thOrder}
\end{figure}

\section{Continuation methods applied to larger size problems}
\label{sec:Problem_g05_100_1}
We now investigate if the insights of the previous section still hold for other and larger problems. To this end, we have performed a continuation analysis of all unweighted graphs of the BiqMac library using the sigmoid nonlinearity of equation \eqref{eq:tanh}. More details on the parameters and stop conditions of the continuation routine used in this section can be found in the methods section \ref{subsec: Continuation routine}. These BiqMac problems have $60$, $80$, or $100$ spins. As we observe qualitatively different behavior for different problems, we will discuss here a number of exemplary bifurcation sequences. All of these problems are Ising hard when using the third order polynomial nonlinearity.

We first take a closer look at problem g05\_100.1 of the BiqMac library and perform a continuation analysis. As was mentioned before, this particular problem becomes Ising easy when solved with the sigmoid nonlinearity and using sufficiently small values of the gain parameter $\alpha$. This changing of difficulty class is shown in figure \ref{fig:tanh_different_alpha}. The results of the continuation analysis are summarised in figure \ref{fig:3Regions_Cusp_g05_100.1_tanh}. The figure shows the tracking of the PB and the two SNs (SN 1 and SN 2) through the parameter space of $\alpha$ and $\beta$. For large $\alpha$-values, two SNs exists, which indicates that the first pitchfork branch is connected to the ground state branch via an unstable branch. The boundaries of this unstable connecting branch are given by the two SNs. The two saddle nodes move closer and closer together as $\alpha$ is decreased, untill they collide in a CP at $\alpha_{\text{CP}}=0.962$. At that point, the two SNs and therefore the unstable branch disappear and the ground state branch is connected directly to the first PB of the origin. This means that the problem is Ising easy when using $\alpha < \alpha_{\text{CP}}$.

\begin{figure}
     \centering
     \includegraphics[width=\linewidth]{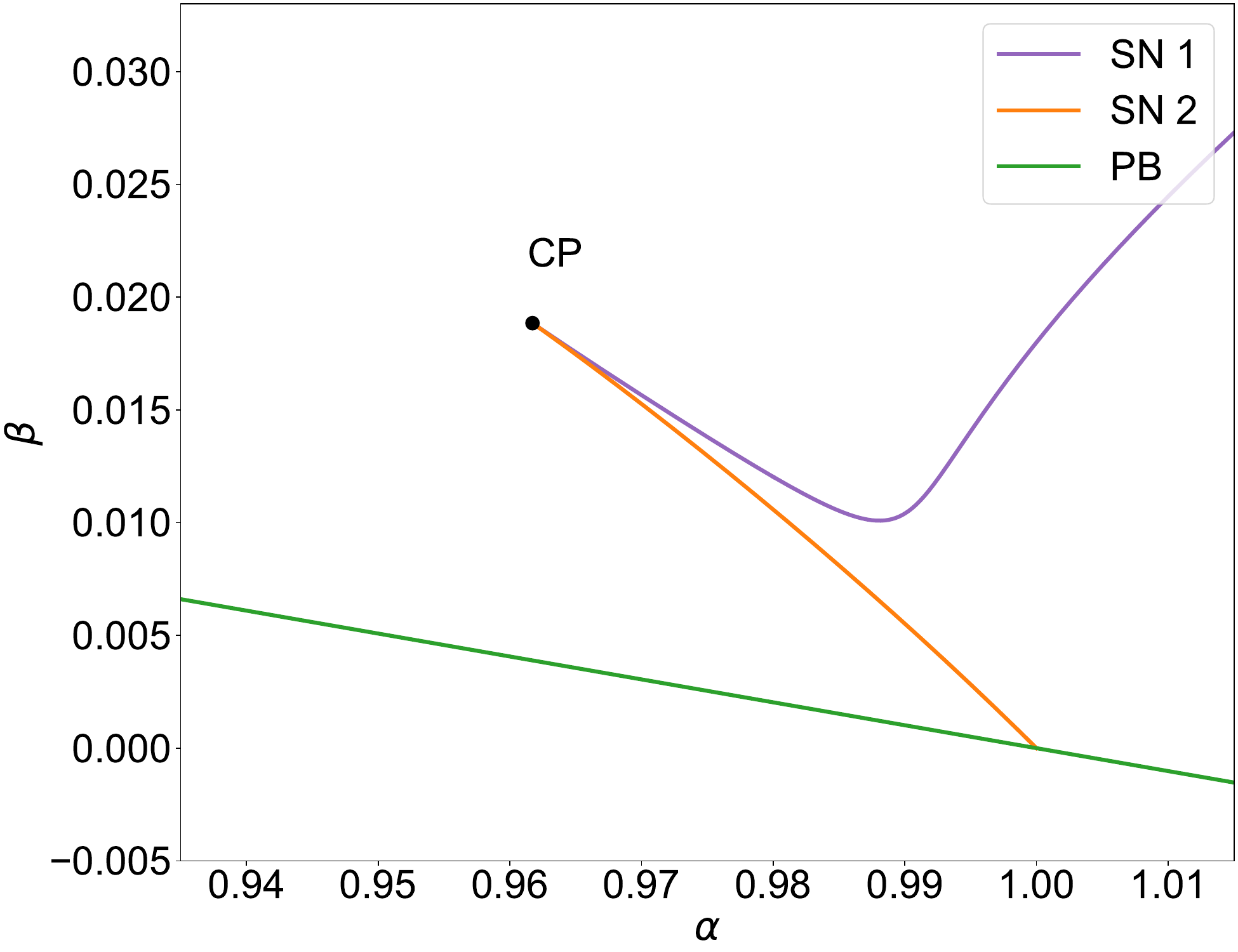}
     \caption{Continuation in the two parameters ($\alpha$ and $\beta$) of the PB and the two SNs for problem g05\_100.1 of the BiqMac library solved with the sigmoid nonlinearity.}
     \label{fig:3Regions_Cusp_g05_100.1_tanh}
\end{figure}

For problem g05\_100.1, discussed in the previous paragraph, the ground state branch and the first pitchfork branch are connected by a single unstable branch. For other problem instances, the first pitchfork branch can be connected to the ground state branch via a multitude of intermediate stable and unstable branches. Problem g05\_100.4 of the BiqMac library is shown as an example in figure \ref{fig:4RegionsCusp_g05_100.4_tanh}. Figure \ref{fig:4RegionsCusp_g05_100.4_tanh}~(a) shows the continuation of the PB and all SNs through the parameter space of $\alpha$ and $\beta$. For low values of the gain $\alpha$, only the PB exists. This indicates that the ground state branch is directly connected to the first PB of the origin as is shown in figure \ref{fig:4RegionsCusp_g05_100.4_tanh}~(b) for $\alpha=0.75$. When $\alpha$ is increased above $\alpha_{\text{CP 1}}=0.803$, a couple of SNs emerge from the CP. The ground state branch is therefore no longer directly connected to the first PB of the origin. Instead, it is connected to the first pitchfork branch via a single unstable branch as shown in figure \ref{fig:4RegionsCusp_g05_100.4_tanh}~(c) for $\alpha=0.815$. In figure \ref{fig:4RegionsCusp_g05_100.4_tanh}~(c), the two SNs that emerged from CP 1 are shown as the orange and brown dots. As the gain $\alpha$ is further increased above $\alpha_{\text{CP 2}}=0.831$, an extra couple of SN emerge, which introduces a new unstable branch. The two new SNs at $\alpha=0.90$ are shown in figure \ref{fig:4RegionsCusp_g05_100.4_tanh}~(d) as the red and purple dots. This results in an intermediate stable branch. When $\alpha$ is increased above $\alpha_{\text{CP 3}}=0.938$, the same process is repeated, which results in an extra unstable branch and intermediate stable branch. This scenario is shown in figure \ref{fig:4RegionsCusp_g05_100.4_tanh}~(e). When $\alpha$ is even further increased above $0.976$, more couples of SN emerge, but these have been omitted for the sake of clarity.

\begin{figure}
     \centering
     \includegraphics[width=\linewidth]{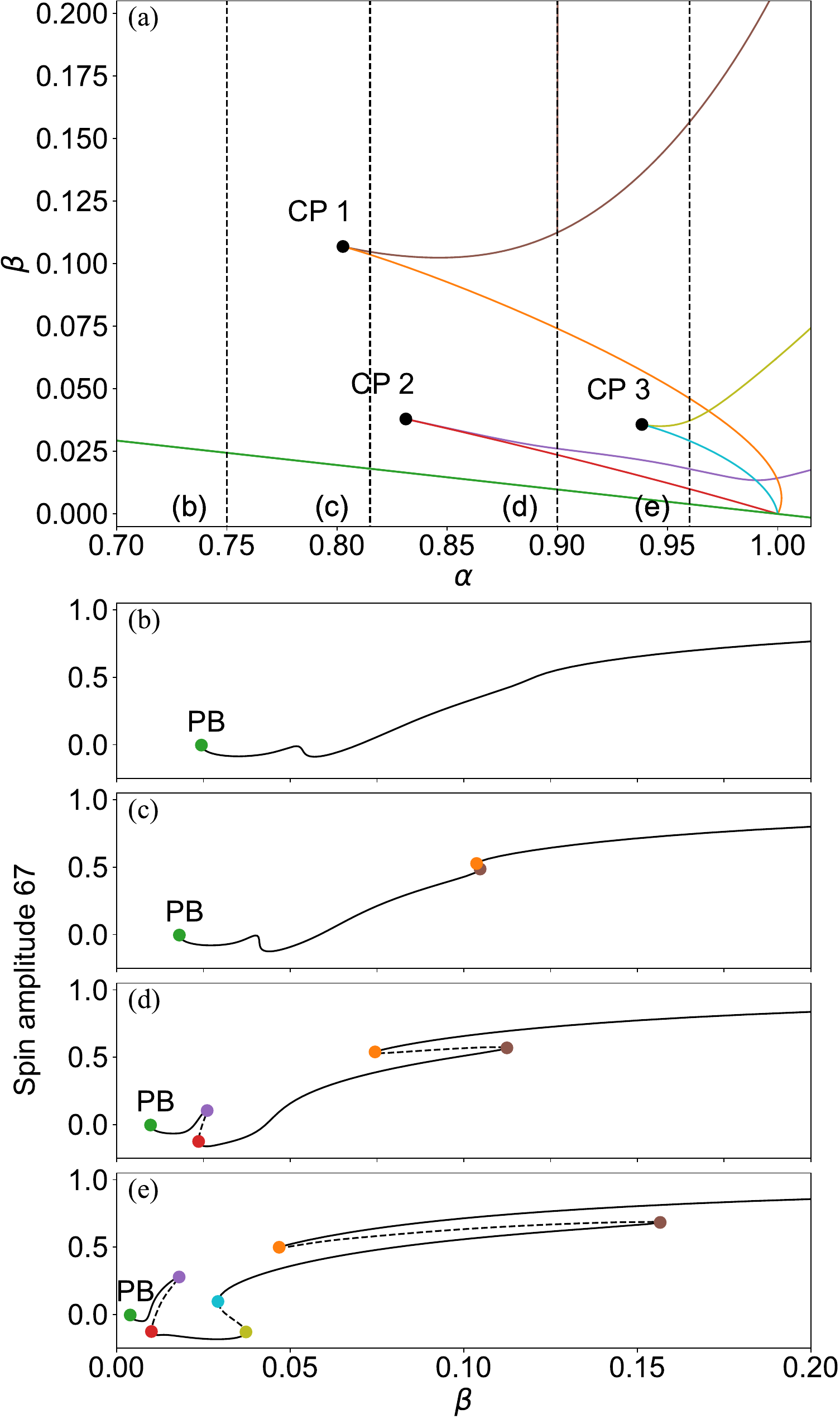}
     \caption{Continuation in the two parameters ($\alpha$ and $\beta$) of the PB and the SNs for problem g05\_100.4 of the BiqMac library solved with the sigmoid nonlinearity.}
     \label{fig:4RegionsCusp_g05_100.4_tanh}
\end{figure}

In the previous two examples, the ground state is connected to the first pitchfork branch either directly or via intermediate branches. For some other problems however, the first pitchfork branch does not connect to the ground state branch. As an example of this case, we show the continuation of the first pitchfork branch of problem g05\_100.3 of the BiqMac library in figure \ref{fig:SpinsAndEnergy_g05_100_3_tanh}. Figure \ref{fig:SpinsAndEnergy_g05_100_3_tanh}~(a) shows all spin amplitudes of the first pitchfork branch as a function of coupling strength $\beta$. It emerges through the PB at $\beta=1.
04\cdot 10^{-4}$ and disappears again through SN 1 in figure \ref{fig:SpinsAndEnergy_g05_100_3_tanh}~(a) at $\beta = 0.51$. The unstable branch of the saddle node bifurcation loops back to lower $\beta$-values and connects to another stable fixed point through SN 2 at $\beta = 0.05$. This is all very similar to problem g05\_100.1 except for the fact that the final state never reaches the ground state energy. This state remains stable until $\beta = 10$ without any changes in the spin amplitudes. After this $\beta$-value, we stopped the continuation algorithm. Figure \ref{fig:SpinsAndEnergy_g05_100_3_tanh}~(b) shows the binary energy of the analog fixed point as a function of coupling strength $\beta$. The ground state energy is shown as a gray dashed line and is never obtained by either of the two branches. This means that although the saddle node bifurcations can be removed by choosing $\alpha$ appropriately, the problem never becomes Ising easy.

%This specific problem also has another particular feature. From figure \ref{fig:SpinsAndEnergy_g05_100_3_tanh} (b), one can see that the energy of the two stable fixed points is the same even though the sign of one spin amplitude is different. It turns out that the $9^\text{th}$ spin amplitude has an equal amount of spin-up neighbors as spin-down neighbors in these fixed points. The sign of the $9^\text{th}$ spin amplitude therefore does not influence the energy of the fixed point and it is this spin amplitude that changes sign on the unstable branch. The unstable branch thus connects two fixed points that are related to each other via this symmetry.

\begin{figure}
     \centering
     \includegraphics[width=\linewidth]{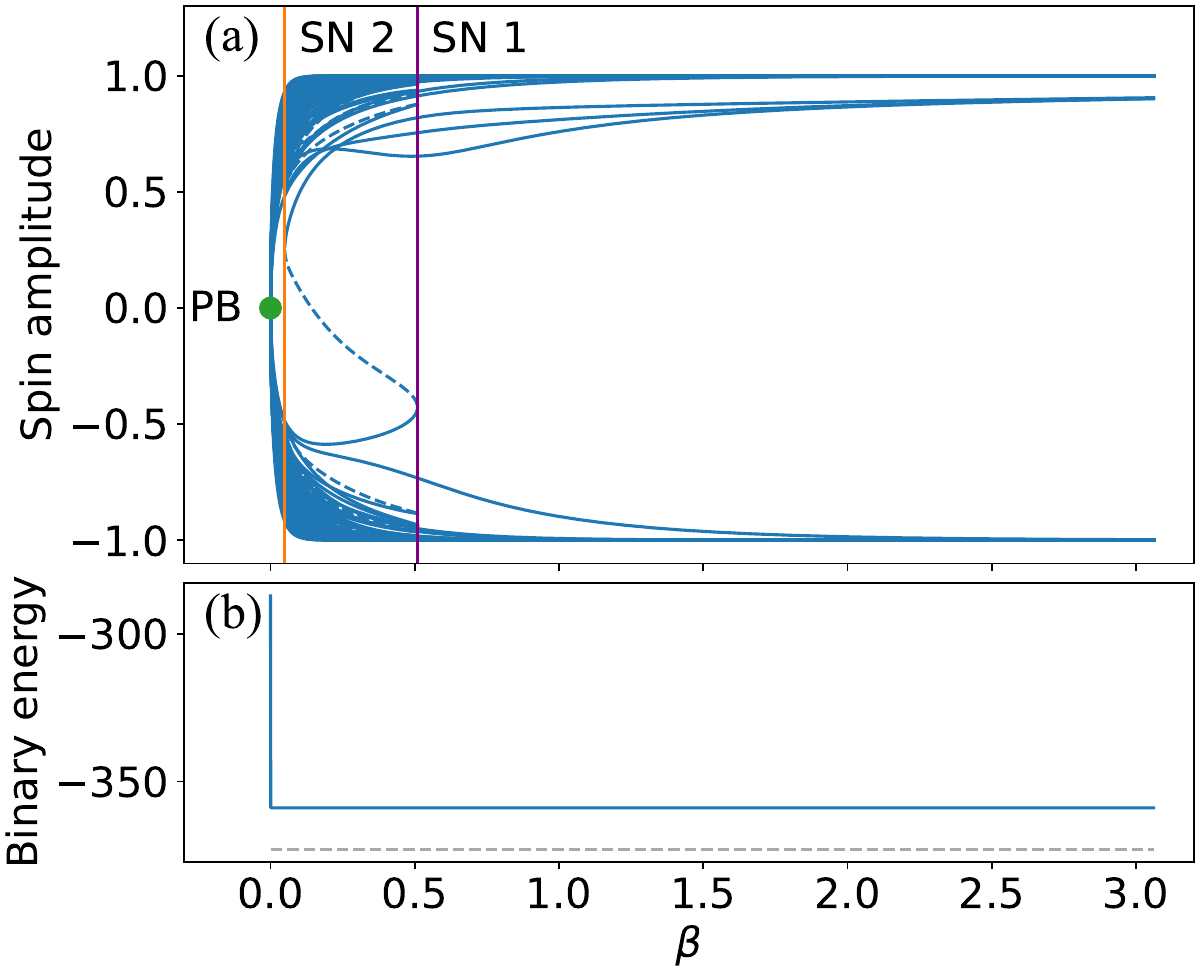}
     \caption{Continuation in coupling strength $\beta$ at $\alpha=0.999$ of problem g05\_100.3 of the BiqMac library when using the sigmoid nonlinearity. (a) The spin amplitudes and (b) the corresponding binary energy. The gray dashed line represents the binary ground state energy.}
     \label{fig:SpinsAndEnergy_g05_100_3_tanh}
\end{figure}

We summarise our results of this continuation analysis on the BiqMac library in table \ref{tab:Ising_Easy_Become}. The first column lists the name of the benchmark problem. The second column lists the $\alpha$-value of the CP below which the problem becomes Ising easy when solved using the sigmoid nonlinearity and in the third column, we list the $\zeta$-value of the CP, above which the problem becomes Ising easy, when using the fifth order polynomial nonlinearity. The case where the ground state branch is not connected to the first PB of the origin, is mentioned in the table as 'Not connected'. If the problem is already Ising easy when using the third order polynomial nonlinearity, the table cell mentions 'ALL' as the problem is then Ising easy for all values of the other free parameter both when using the sigmoid and when using the fifth order polynomial nonlinearity.

%Instance g05\_100.0 has a degenerate ground state. Similar to what was observed for problem g05\_100.3, the fixed point originating at the first pitchfork bifurcation of the origin is connected to another state with the same binary energy via an unstable branch. In this case however, both these states have the ground state energy and therefore correspond to two different ground state configurations. This problem is therefore always Ising easy, but the $\alpha$-value mentioned in the third column of Table \ref{tab:Ising_Easy_Become} is the value below which the unstable branch disappears and the first of the degenerate ground states continuously evolves to the second.

We see that out of thirty problem instances, only five are Ising easy when using the third order polynomial nonlinearity. This amount increases to eight when using the fifth order nonlinearity and to fifteen when using the sigmoid nonlinearity if the other free parameter is chosen appropriately. So the amount of problems that can be tackled using annealing schemes is roughly tripled when using the sigmoid nonlinearity compared to the case when the third order polynomial nonlinearity is used. Moreover, this table is also useful to identify problems that are Ising hard and which are therefore particularly suited to benchmark different Ising machine algorithms and Ising machines. 

\begin{table}
\caption{List of $\alpha$ ($\zeta$)-values below (above) which the problem becomes Ising easy when solved using the sigmoid (fifth order polynomial) nonlinearity.}
    {
    \centering
    \begin{tabular}{|p{0.20\linewidth}|p{0.32\linewidth}|p{0.32\linewidth}|}
    \hline
    \begin{center}
        Problem name
    \end{center}
     & \begin{center}
         $\alpha_{\text{CP}}$\\ for sigmoid
     \end{center} & \begin{center}
         $\zeta_{\text{CP}}$\\ for fifth order
     \end{center}\\
    \hline\hline
    \vspace{0.001cm}
    g05\_60.0 & \vspace{0.001cm}0.803&\vspace{0.001cm}0.133\\
    g05\_60.1 & Ising easy&Ising easy\\
    g05\_60.2   & \multicolumn{1}{|r|}{Not connected}&\multicolumn{1}{|r|}{Not connected}\\
    g05\_60.3   & \multicolumn{1}{|r|}{Not connected}&\multicolumn{1}{|r|}{Not connected}\\
    g05\_60.4   & 0.887&\multicolumn{1}{|r|}{Not connected}\\
    g05\_60.5   & 0.361&\multicolumn{1}{|r|}{Not connected}\\
    g05\_60.6   & \multicolumn{1}{|r|}{Not connected}&\multicolumn{1}{|r|}{Not connected}\\
    g05\_60.7   & 0.583&\multicolumn{1}{|r|}{Not connected}\\
    g05\_60.8   & \multicolumn{1}{|r|}{Not connected}&\multicolumn{1}{|r|}{Not connected}\\
    g05\_60.9   & \multicolumn{1}{|r|}{Not connected}&\multicolumn{1}{|r|}{Not connected}\\
    \hline
    g05\_80.0   & \multicolumn{1}{|r|}{Not connected}&\multicolumn{1}{|r|}{Not connected}\\
    g05\_80.1 & Ising easy&Ising easy\\
    g05\_80.2   & 0.835&\multicolumn{1}{|r|}{Not connected}\\
    g05\_80.3   & \multicolumn{1}{|r|}{Not connected}&\multicolumn{1}{|r|}{Not connected}\\
    g05\_80.4   & 0.867&0.040\\
    g05\_80.5   & \multicolumn{1}{|r|}{Not connected}&\multicolumn{1}{|r|}{Not connected}\\
    g05\_80.6   & 0.903&\multicolumn{1}{|r|}{Not connected}\\
    g05\_80.7   & 0.739&\multicolumn{1}{|r|}{Not connected}\\
    g05\_80.8   & \multicolumn{1}{|r|}{Not connected}&\multicolumn{1}{|r|}{Not connected}\\
    g05\_80.9   & \multicolumn{1}{|r|}{Not connected}&\multicolumn{1}{|r|}{Not connected}\\
    \hline
    g05\_100.0 & Ising easy&Ising easy\\
    g05\_100.1   & 0.962&0.104\\
    g05\_100.2 & Ising easy&Ising easy\\
    g05\_100.3   & \multicolumn{1}{|r|}{Not connected}&\multicolumn{1}{|r|}{Not connected}\\
    g05\_100.4   & 0.803&\multicolumn{1}{|r|}{Not connected}\\
    g05\_100.5   & \multicolumn{1}{|r|}{Not connected}&\multicolumn{1}{|r|}{Not connected}\\
    g05\_100.6   & Connected, but never Ising easy &\multicolumn{1}{|r|}{Not connected}\\
    g05\_100.7   & \multicolumn{1}{|r|}{Not connected}&\multicolumn{1}{|r|}{Not connected}\\
    g05\_100.8   &\multicolumn{1}{|r|}{Not connected} &\multicolumn{1}{|r|}{Not connected}\\
    g05\_100.9 & Ising easy&Ising easy\\
    \hline
    \end{tabular}\par
    }
    \label{tab:Ising_Easy_Become}
\end{table}

%\begin{table}
%    {
%    \centering
%    \begin{tabular}{|p{0.25\linewidth}|p{0.28\linewidth}|p{0.28\linewidth}|}
%    \hline
%    Problem name & Ising easy with third order nonlinearity & $\alpha$-value below which the problem becomes Ising easy when using the sigmoid nonlinearity\\
%    \hline\hline
%    g05\_60.0 && 0.803\\
%    g05\_60.1 &YES& ALL\\
%    g05\_60.2 && \\
%    g05\_60.3 && \\
%    g05\_60.4 && 0.887\\
%    g05\_60.5 && 0.361\\
%    g05\_60.6 && \\
%    g05\_60.7 && 0.583\\
%    g05\_60.8 && \\
%    g05\_60.9 && \\
%    \hline
%    g05\_80.0 && \\
%    g05\_80.1 &YES& ALL\\
%    g05\_80.2 && 0.835\\
%    g05\_80.3 && \\
%    g05\_80.4 && 0.867\\
%    g05\_80.5 && \\
%    g05\_80.6 && 0.903\\
%    g05\_80.7 && 0.739\\
%    g05\_80.8 && \\
%    g05\_80.9 && \\
%    \hline
%    g05\_100.0* & YES& 0.924 (ALL)\\
%    g05\_100.1 && 0.962\\
%    g05\_100.2 & YES& ALL\\
%    g05\_100.3 && \\
%    g05\_100.4 && 0.803\\
%    g05\_100.5 && \\
%    g05\_100.6 && \\
%    g05\_100.7 && \\
%    g05\_100.8 && \\
%    g05\_100.9 &YES& ALL\\
%    \hline
%    \end{tabular}\par
%    }
%    \caption{List of $\alpha$-values below which the problem becomes Ising easy when solved using the sigmoid nonlinearity.}
%    \label{tab:Ising_Easy_Become}
%\end{table}

\section{Discussion}
Each physical implementation of Ising machines comes with its own nonlinearity in the evolution of the spin amplitudes. The fifth order polynomial nonlinearity can be seen as a general expression for the first five orders of the Taylor expansion of that nonlinearity. As the fifth order polynomial nonlinearity is able to render some problems Ising easy, it stands to reason that any nonlinearity that requires at least a fifth order term in its Taylor expansion with at least two independent parameters has that capability. However, from the comparison between the sigmoid and the fifth order polynomial nonlinearity in table \ref{tab:Ising_Easy_Become}, it is clear that the former outperforms the latter in the number of problems they can render Ising easy. We speculate that the reason for this stems from the saturation of spin amplitudes when utilising the sigmoid nonlinearity. This saturation results in a phasespace that is confined to a hypercube ($x_i\in[-1,1]$) and would mean that as $\beta$ increases, more and more fixed points are cramped in the same hypervolume. This will likely increase the probability that the ground state branch is connected to the first pitchfork branch. Furthermore, we observe that using the sigmoid nonlinearity results in less amplitude inhomogeneity, which has been observed to improve performance \cite{Order_of_magnitude, CombinatorialOptimization_Leleu}. Nevertheless, we can conclude that the capability of the nonlinearity in turning problems Ising easy is an interesting characteristic that should not be ignored when considering a possible physical implementation of an Ising machine.

It is clear that the nonlinearity used in the dynamical evolution of the spin amplitudes has a huge impact on the performance of Ising machines. So having a way to implement different nonlinearities can be very useful. There are already technologies out there that can provide a reconfigurable nonlinear element that could potentially be implemented in already existing set-ups of Ising machines \cite{AllOpticalNonlinearActivationFunction, Electro-OpticNonlinearActivationFunction, NonlinearPhotonicActivationFunction}.

It is also worth noting that although continuation methods are used here to investigate fixed points of problems solved by Ising machines, these methods can also be used as solvers of Ising networks. They are a way to consistently track the first pitchfork branch and they even manage to move to connected branches via the unstable connecting branches. This means that continuation methods can be used to obtain the ground state of all Spectral and Ising easy problems as well as the Ising hard problems that can be rendered Ising easy. The advantage of continuation methods is that they can be used without having to choose the gain $\alpha$ appropriately as they are able to follow the unstable connecting branches. Furthermore, there are other Ising hard problems that can be solved rather easy with these continuation methods, while they are difficult to tackle with annealing schemes. This is generally the case when the first pitchfork branch is connected to the ground state branch via a series of unstable and stable branches. But at least one of those unstable branches does not disappear for any combination of the parameters, which means that these problems never become Ising easy. An example of such a problem is instance g05\_100.6. In this case annealing schemes cannot be utilised effectively, but continuation methods would have no problem obtaining the ground state.

In this study, we thus show that continuation methods are an indispensable tool to study Ising machines: they show which problems will be hard to solve, they explain and predict which Ising machine implementations work well and they offer a new algorithmic approach to obtain the ground state.

\section{Methods}
\label{sec:Methods}

\subsection{Simulating the temporal evolution of the Ising machine}
\label{subsec:SimulationOfTheIsingMachine}
A simulation of the Ising machine using the third order polynomial nonlinearity consists of integrating the transfer function given by equation \eqref{eq:ThirdOrder}. This is done using the Euler integration method. First, the spin amplitudes are initialised by drawing $N$ random numbers (one for each analog spin) from a Gaussian distribution with zero mean and a standard deviation of $0.001$. Then, all spins are updated according to the following rule
\begin{multline}
    x_i(t+1) = x_i(t) \\+ \frac{h}{\tau}\left((\alpha - 1)x_i(t) - x_i^3(t) + \beta(t)\sum_{j}J_{ij}x_j(t)\right),
\end{multline}
where $h$ is the time step size and $\tau$ is the intrinsic time-scale of the Ising machine. The time step size used here is $h=0.01$ and the units of time are chosen such that $\tau = 1$. The gain $\alpha$ and the coupling strength $\beta$ are fixed during the entire run. 

This updating procedure is repeated $T$ times, where $T$ is the run time parameter. The run time should be chosen sufficiently large so that all spin amplitudes reach a stable value and are not changed significantly by the update rule. The binary spins corresponding to the spin amplitudes $x_i$ of the Ising machine can be obtained at any time by performing the mapping of equation \eqref{eq:map}. Using these binary spins, the energy can be obtained using equation \eqref{eq:Hamiltonian}. In this article, we focus on the benchmark problems of the BiqMac library because for these problems, the ground state energy is known. Therefore, the resulting energy of the Ising machine can be compared with the known ground state energy to check whether the Ising machine reached the ground state. The success rate is defined as the fraction of the runs that reached this ground state.

The exact value of the parameters $\alpha$ and $\beta$ heavily influences the chance of reaching the ground state \cite{Order_of_magnitude}. Therefore, to determine the optimal values of $\alpha$ and $\beta$, a scan is performed. The Ising machine is initialised $100$ times for $450$ different combinations of $\alpha$ and $\beta$ with $-1 < \alpha < 1$ and $0.1 < \beta < 1$. The combination of parameters $\alpha$ and $\beta$ for which the ground state is most often reached after $5000$ time steps is taken as the optimal parameters for that problem. This scan was performed for each problem that was solved. 

\subsection{Annealing scheme}
\label{subsec:Annealing_scheme}
The annealing scheme used in this article is done using the following steps. We first set the system parameters such that we start at the first pitchfork bifurcation of the origin. In supplementary material note \ref{subsec:Bifurcation_points_origin}, it is shown that the first pitchfork bifurcation of the origin occurs at
\begin{equation}
    \label{eq:First_Bifurcation_condition}
    \beta^*_1 = \frac{1-\alpha}{\mu_1},
\end{equation}
where $\mu_1$ is the largest eigenvalue of the coupling matrix. Next, we set all spin amplitudes to random values close to zero. We then integrate equation \eqref{eq:ThirdOrder} using a regular Euler integration method, but slightly increase the coupling strength parameter $\beta$ to $\beta + \beta_{\text{step}}$ at each time step. When $\beta>\beta^*_1$, the origin will become unstable and the system will get pushed to one of the first pitchfork branches. For Ising easy problems, this fixed point does not yet correspond to the ground state of the binary system when the mapping in equation \eqref{eq:map} is applied. So, we continue to increase $\beta$ slightly at every time step of the simulation until this fixed point eventually evolves to represent the ground state of the binary system. It is important to note that altering $\beta$ will change the energy landscape. This means that $\beta$ has to be altered slow enough so that the system relaxes to the fixed point at each time step. For a general problem, the ground state energy is not a priori known. This means that one cannot see from the binary energy if $\beta$ is large enough for the steady state to get mapped to the binary ground state or if $\beta$ has to be annealed further to higher values. However, if $\beta$ is large enough, the fixed point gets mapped to a minimum of the binary system and the following condition is fulfilled (see section \ref{subsec:Sensible_mapping}):
\begin{equation}
    \label{eq:CorrectMapping_text}
    \text{sign}\left(\sum_j J_{ij}\frac{x^*_j}{x^*_i}\right) = \text{sign}\left(\sum_j J_{ij} \sigma_i\sigma_j\right),
\end{equation}
where $x_i^*$ is the spin amplitude when the system is in a fixed point. So far, we have not observed this condition being fulfilled for Ising easy problems without simultaneously having reached the binary ground state. So, checking if this condition is fulfilled for all $i$ allows the  endpoint of the annealing scheme to be determined during the run. Note that this annealing scheme will also solve spectral easy problems.

We however notice that for some problems that are not spectral or Ising easy, this condition is never fulfilled for the first pitchfork branch. So, when the difficulty class of a problem is not known, it might be useful to add a second stopping condition to avoid infinite loops. One such a condition could be that the absolute value of the smallest spin amplitude should be higher than some threshold. Because if that is the case, more spin flips are less likely to occur.

\subsection{Continuation routine}
\label{subsec: Continuation routine}
The continuation analysis in this article is done using the software ‘auto-07p’ \cite{AUTO}. We track the origin as the coupling strength $\beta$ is increased, until the first PB and then switch to the first pitchfork branch. If this branch would disappear through a SN, we track the unstable branch to lower $\beta$-values until it connects to a stable branch through a second SN. 

When using the sigmoid nonlinearity in equation \eqref{eq:tanh}, we continue this tracking until either we obtain the ground state energy or all spin amplitudes saturate at $\pm1$. This saturation condition can be used as the spin amplitudes cannot obtain absolute values larger than $1$ when using the sigmoid nonlinearity. The continuation was done with a gain $\alpha=0.999$. This value is chosen as we want to observe the SNs and unstable branches, which gradually disappear as $\alpha$ is lowered from $1$. 

Most continuations using the fifth order nonlinearity are performed using $\zeta=0.1$. However, sometimes the software got stuck and for those instances (g05\_80.8, g05\_100.4 and g05\_100.8), the arbitrarily chosen $\zeta=0.5$ was used instead. As the spin amplitudes can obtain absolute values larger than $1$ for large values of $\beta$ when using the fifth order polynomial nonlinearity, the stop condition is different from the case of the sigmoid nonlinearity. In this case, when the first pitchfork branch does not seem to obtain the ground state energy or disappear through a SN, we stop the continuation at $\beta=50$. We assume that after this $\beta$-value, the state will not bifurcate or change its energy anymore. 

When the ground state branch of a specific problem instance bifurcates through a SN, but is connected to the first PB of the origin via one or more unstable branches, we then track the SNs for different values of $\alpha$ and $\beta$. A general example is shown in figure \ref{fig:4RegionsCusp_g05_100.4_tanh}. We continue this tracking until all CPs are found. After that, we track the origin to the first pitchfork branch for a slightly lower $\alpha$-value than $\alpha_{\text{CP}}$ to check if now the ground state branch is directly connected to the first PB of the origin.

\section{Data availability}
The authors declare that all relevant data are included in the manuscript. Additional data are available from the corresponding author upon reasonable request.

\section{Author contributions}
J.L. performed the simulations and wrote the manuscript. G.V. and G.V.d.S. supervised the project. All authors discussed the results and reviewed the manuscript.

\section{Additional information}
{\bf Competing interests:} The authors declare no competing interests.\\
{\bf Acknowledgements:} The authors would like to thank Thomas Van Vaerenbergh and Fabian Böhm for the insightful discussions. \\This research was funded by the Research Foundation Flanders (FWO) under grants G028618N, G029519N and G006020N. Additional funding was provided by the EOS project "Photonic Ising Machines". This project (EOS number 40007536) has received funding from the FWO and F.R.S.-FNRS under the Excellence of Science (EOS) programme.

\clearpage


\clearpage
\section{Supplementary material}
\subsection{Rescaling of the analog spin dynamics}
\label{subsec:Renormalisation}
The time evolution of the spin amplitudes when using the the third order polynomial nonlinearity is given by
\begin{equation}
    \label{eq:AppendixThirdOrder}
    \frac{dx_i}{dt} = (\alpha-1)x_i - x_i^3 + \beta\sum_j^N J_{ij}x_j.
\end{equation}
Assuming $\alpha < 1$, rescaling the time $t' = (1-\alpha)t$ results in 
\begin{equation}
    \frac{dx_i}{dt'} = -x_i - \frac{1}{1-\alpha} x_i^3 + \frac{\beta}{1-\alpha}\sum_j^N J_{ij}x_j.
\end{equation}
Rescaling the spin amplitude $x_i' = x_i\sqrt{\frac{1}{1-\alpha}}$ and redefining $\beta' = \frac{\beta}{1-\alpha}$ gives
\begin{equation}
    \label{eq:AppendixRescaledThirdOrder}
    \frac{dx'_i}{dt'} = -x'_i - (x'_i)^3 + \beta'\sum_j^N J_{ij}x'_j.
\end{equation}
From this, it is clear that only the fraction of the gain $\alpha$ and the coupling strength $\beta$ is important. Using equation \eqref{eq:AppendixRescaledThirdOrder} instead of equation \eqref{eq:AppendixThirdOrder} effectively removes a parameter from the model.

A similar rescaling can be done for the time evolution using the fifth order polynomial nonlinearity to remove the gain parameter $\alpha$ from the model. This model therefore only has two independent free parameters: $\beta$ and $\zeta$.

Such a rescaling cannot be done for the dynamical equations using the sigmoid nonlinearity. this model therefore has an extra parameter.

\subsection{Using annealing schemes to tackle Ising easy problems}
\label{sec:Annealing}
In order to test the effectiveness of the annealing scheme from section \ref{subsec:Annealing_scheme}, we use it to solve all Ising easy problems of the BiqMac library that are unweighted and have an edge density of 50\% using the third order polynomial nonlinearity of equation \eqref{eq:ThirdOrder}. We solve each instance $1000$ times and compare the success rate of finding the ground state energy using this annealing scheme with the success rate of the more commonly applied approach. In this last approach, we set the coupling strength $\beta$ well above the $\beta$-value of the first pitchfork bifurcation of the origin, initialise the spin's amplitude to random values close to zero and let the Ising machine evolve until a steady state is reached. More details can be found in the methods section \ref{subsec:SimulationOfTheIsingMachine}. The success rate for problems when no annealing scheme is utilised is shown in the second column of table \ref{tab:Annealing_scheme}. The third and fourth columns show the success rate when the proposed annealing scheme is used for two values of 
$\beta_{\text{step}}$. To have a fair comparison, the simulations without annealing are performed with roughly the same amount of time steps as the simulation with annealing and a $\beta_{\text{step}}=10^{-5}$. We observe an increase in success rate for all problems when the annealing scheme is used. Furthermore, we see that when the step size of $\beta$ is decreased further, the success rate becomes $100$\% for all problems. Note however that these are the results of simulations without noise. These are meant as a proof of concept that annealing schemes can be used efficiently to tackle Ising easy problems. We expect that when using a physical Ising machine to perform these annealing schemes the success rate will be lower due to noise.

\begin{table}[h!]
\caption{Comparison of the success rate of the Ising machine with and without annealing scheme for several Ising easy problems.}
    {
    \centering
    \begin{tabular}{|p{0.17\linewidth}|p{0.2\linewidth} |p{0.2\linewidth}|p{0.2\linewidth}|}
    \hline
    Problem name & Success rate without annealing (\%) & Success rate with annealing ($\beta_{\text{step}}=10^{-5}$) (\%) & Success rate with annealing ($\beta_{\text{step}}=10^{-6}$) (\%)\\
    \hline\hline
    g05\_60.1 & 80.8 & 99.8 & 100\\
    g05\_80.1 & 100 & 100 & 100\\
    g05\_100.0 & 31.7 & 97.6 & 100\\
    g05\_100.2 & 81 & 100 & 100\\
    g05\_100.9 & 46.2 & 96.7 & 100\\
    \hline
    \end{tabular}\par
    }
    \label{tab:Annealing_scheme}
\end{table}

\subsection{Bifurcation points of the origin}
\label{subsec:Bifurcation_points_origin}
The eigenvalues of the Jacobian matrix evaluated at a fixed point can be used to determine the stability of that fixed point. When an eigenvalue is negative (positive), the fixed point is attracting (repellent), along the direction determined by the eigenvector corresponding to that eigenvalue. Therefore, when all eigenvalues are positive, the fixed point is said to be stable. When the largest eigenvalue of the Jacobian matrix evaluated at a fixed point changes from a negative number to a positive one, that fixed point becomes unstable. This corresponds to a bifurcation. One can therefore know the pitchfork bifurcation points of the origin by calculating the largest eigenvalue of the Jacobian matrix at the origin.
The Jacobian matrix of the third order polynomial transfer function of equation \eqref{eq:ThirdOrder} is given by
\begin{equation}
    \label{eq:Jacobian_ThirdOrder}
    \mathcal{U}_{ij} = (\alpha-1)\delta_{ij}-3x_i^2\delta_{ij} + \beta J_{ij}
\end{equation}
and for the sigmoid transfer function, the Jacobian becomes
\begin{multline}
    \label{eq:Jacobian_Sigmoid}
    \mathcal{U}_{ij} = -\delta_{ij} \\+ \left(1 - \tanh^2\left(\alpha x_i + \beta \sum_{j}^N J_{ij} x_j\right)\right)(\alpha \delta_{ij} + \beta J_{ij}).
\end{multline}
When evaluated at the origin, both reduce to
\begin{equation}
    \label{eq:Jacobian_origin}
    \mathcal{U}_{ij}|_{\vec{x}=\vec{0}} = (\alpha-1)\delta_{ij} + \beta J_{ij}.
\end{equation}
The eigenvalues of this matrix $\lambda_i$ are given by
\begin{equation}
    \label{eq:Eigenvalues_Jacobian}
    \lambda_i = \alpha - 1 + \beta \mu_i,
\end{equation}
where $\mu_i$ are the eigenvalues of the coupling matrix $J_{ij}$. Every time an eigenvalue of the Jacobian matrix changes sign, a bifurcation happens. This occurs for the critical values $\beta^*$ given by
\begin{equation}
    \label{eq:Bifurcation_condition}
    \beta^*_i = \frac{1-\alpha}{\mu_i^+},
\end{equation}
where $\mu_i^+$ is the $i^{\text{th}}$ positive eigenvalue of the coupling matrix. So for a given value of $\alpha$, the bifurcation points of the origin can be easily found by calculating the eigenvalues of the coupling matrix.

One can easily show that the bifurcation points for the rescaled fifth order polynomial nonlinearity of equation \eqref{eq:FifthOrder} are given by
\begin{equation}
    \label{eq:Bifurcation_condition_without_alpha}
    \beta^*_i = \frac{1}{\mu_i^+}.
\end{equation}

\subsection{Sensible mapping to the binary system}
\label{subsec:Sensible_mapping}
Solving the Ising problem is achieved by finding the state $\vec{\sigma}$, consisting of $N$ spins $\sigma_i$, that minimizes the Ising hamiltonian:
\begin{equation}
    \label{eq:hamiltonian}
    \mathcal{H}_{\text{Ising}} = -\frac{1}{2}\vec{\sigma}^T J \vec{\sigma} = -\sum_{i<j} J_{ij} \sigma_i\sigma_j,
\end{equation}
where $\sigma_i \in \{0, 1\}$ and $J$ is the coupling matrix. %This is often taken to be a symmetric matrix with diagonal entries equal to zero. 
Starting from any state, flipping a spin will alter the energy by 
\begin{align*}
    (\Delta E)_i &= \left(-\sum_j J_{ij} (-\sigma_i)\sigma_j\right) - \left(-\sum_j J_{ij} \sigma_i\sigma_j\right)\\
    &= 2\sum_j J_{ij} \sigma_i\sigma_j.
\end{align*}
A state $\Vec{\sigma}$ is said to be a minimum if flipping any spin increases the energy, i.e. if for all i,
\begin{equation}
    \label{eq:local_minimum_binary_system}
    \sum_j J_{ij} \sigma_i\sigma_j \geqslant 0.
\end{equation}
When using a third order polynomial as the nonlinearity, the Ising machine tries to solve the Ising problem by dynamically evolving spin amplitudes $\vec{x}$ using the following set of nested differential equations:
\begin{equation}
    \label{eq:dynamics}
    \frac{dx_i}{dt} = (\alpha -1)x_i - x_i^3 + \beta\sum_j J_{ij}x_j,
\end{equation}
where $x_i \in \mathbb{R}$ and $\alpha$ and $\beta$ are the gain and the coupling strength respectively. When the system settles into a fixed point $\vec{x}^*$, the binary solution is obtained by the map $\sigma_i = \text{sign}(x^*_i)$. Of course, the goal is that these fixed points get mapped to minima of the binary system. Due to the analog nature of the spin amplitudes, however, this is not necessarily the case. In the following, we derive a condition that determines if the mapping of the spin amplitudes to binary spins results in a minimum of the binary system. 

The fixed points of the analog system $\Vec{x}^*$ are defined as the points where equation \eqref{eq:dynamics} is equal to zero for all spins.
\begin{gather*}
    (\alpha -1)x^*_i - x^{*3}_i + \beta\sum_j J_{ij}x^*_j = 0\\
    (1 - \alpha) + x^{*2}_i = \beta\sum_j J_{ij}\frac{x^*_j}{x^*_i}\\
    \sum_j J_{ij}\frac{|x^*_j|}{|x_i^*|}\sigma_i\sigma_j > 0,
\end{gather*}
where in the last line, we assumed $\alpha < 1$. If the spin amplitudes are homogeneous, i.e. $|x^*_i| = |x^*_j|\;\forall i, j$, this condition becomes the condition of a minimum of the binary system in equation \eqref{eq:local_minimum_binary_system}. This means that taking the sign of a homogeneous fixed point will result in a minimum of the binary Ising network. If the spin amplitudes are inhomogeneous however, the coupling matrix gets warped:
\begin{equation*}
    J_{ij}^{\text{eff}} = J_{ij}\frac{|x^*_j|}{|x^*_i|}.
\end{equation*}
Because of this, a fixed point does not necessarily get mapped to a minimum of the binary system. It only does when, for all $i$,
\begin{equation}
    \label{eq:CorrectMapping}
    \text{sign}\left(\sum_j J_{ij}\frac{x^*_j}{x^*_i}\right) = \text{sign}\left(\sum_j J_{ij} \sigma_i\sigma_j\right) > 0.
\end{equation}
This is a condition that can be checked during a simulation and that depends on the spin amplitudes, which in turn depend on the system parameters (in this case $\beta)$. So, when an Ising machine settles into a fixed point, the state should be annealed to higher values of $\beta$ until the above condition is fulfilled.

%\newpage
%\appendix
%\section{Difficulty classification of the BiqMac library}
%\begin{table}[h!]
%    {
%    \centering
%    \begin{tabular}{|p{0.2\linewidth} |p{0.2\linewidth}||p{0.2\linewidth} |p{0.2\linewidth}|}
%    \hline
%    Problem name & Difficulty class & Problem name & Difficulty class \\
%    \hline\hline
%    g05\_60.0 & H & g05\_80.0 & H\\
%    g05\_60.1 & E & g05\_80.1 & E\\
%    g05\_60.2 & H & g05\_80.2 & H\\
%    g05\_60.3 & H & g05\_80.3 & H\\
%    g05\_60.4 & H & g05\_80.4 & H\\
%    g05\_60.5 & H & g05\_80.5 & H\\
%    g05\_60.6 & H & g05\_80.6 & H\\
%    g05\_60.7 & H & g05\_80.7 & H\\
%    g05\_60.8 & H & g05\_80.8 & H\\
%    g05\_60.9 & H & g05\_80.9 & H\\ 
%    \hline
%    \end{tabular}\par
%    }
%    \caption{Difficulty classification of the BiqMac library. E means the problem is Ising easy, while H means the problem is Ising hard when using the third order polynomial nonlinearity.}
%    \label{tab:Difficulty_Classes_BiqMac}
%\end{table}

\end{document}